\documentclass[aps,twocolumn,nofootinbib,longbibliography]{revtex4-1}

\usepackage{geometry}                		
\usepackage{graphicx}				
\def\be{\begin{equation}}
\def\ee{\end{equation}}
\usepackage{amssymb}
\begin{document}

\title{{\Large LQG predicts the Unruh Effect}\\[2mm]
Comment to the paper ``Absence of Unruh effect in polymer quantization"
by Hossain and 
Sardar}
\author{Carlo Rovelli}
\affiliation{\vspace{1mm}\mbox{CPT, Aix-Marseille Universit\'e, Universit\'e de Toulon, CNRS,}\\ 
\mbox{Samy Maroun Center for Time, Space and the Quantum.}\\ 
Case 907, F-13288 Marseille, France.}
\date{\small\today}

\begin{abstract}
\noindent 
A  recent  paper claims that loop quantum gravity predicts the absence of 
the Unruh effect.   I show that this is not the case, and take advantage of this opportunity to shed some light on some related issues.

\end{abstract}

\maketitle


\maketitle

The Unruh effect is the fact that a standard thermometer moving at constant acceleration $a$ in the vacuum state of a quantum field on flat space measures a temperature $T=a\hbar/2\pi$, in units where $c=k_{Boltzaman}=1$.   The paper \cite{Hossain2014} claims that Loop Quantum Gravity (LQG) predicts the absence of this effect as a consequence of the fact that LQG modifies a quantum field theory at high-energy (short distance).  The claim is wrong, because the Unruh effect is a low-energy (large distance) phenomenon, not a high-energy (short distance) one.  An accelerated local detector interacting with a quantum field field tests the properties of the field only at length scales of order $L\sim 1/a$. The phenomenon is not sensible to the behaviour of the field at shorter scales.  

In fact, there are a number of derivations of the Unruh effect, and some of these involve \emph{only} the low-energy sector of the theory.  Other derivations, like Unruh's original one, go though the high-energy sector in a convoluted manner, and require an infinite renormalization.  The need for an infinite subtraction is a weakness of the derivation method, not a feature of the phenomenon itself, which can be derived in a straightforward manner.  Below, for completeness, I recall one manner to derive the Unruh effect which is explicitly insensible to the short-scale behavior of the field.  

In the paper \cite{Hossain2014}, the authors recognise that LQG modifies only the short-distance behaviour of the field, not its large-distance behaviour.  This implies that the prediction of the Unruh effect cannot be modified, at least until the accelerations reaches the scale where quantum gravitational effects become relevant, namely the Planck acceleration $a\sim1/ \sqrt{\hbar G}$, a scale where, as argued in \cite{Rovelli2013d}, it is acceleration itself to be bound by LQG effects.  So, I take $a \ll1/ \sqrt{\hbar G}$ below. 

I give here a brief derivation of the Unruh Effect from LQG (in part following \cite{Bianchi2012a}, see also \cite{Rovelli}). A thermometer can be modelled by a two-state system with an energy gap $\epsilon$ between its two energy eigenstates $(|0\rangle,|1\rangle)$, interacting with a generic self-adjoint observable $A$ of a system via a simple interaction term such as
\be
V(t)=g\ (|0\rangle\langle 1|+|1\rangle\langle 0|)\ A(t).
\ee
where $g$ is a small coupling constant. The amplitude for the thermometer to have moven up at time $t$, with the system going from an initial time independent state $|i\rangle$ to a final state $|f\rangle$ is given to first order in $g$ by Fermi's golden rule:
\begin{eqnarray}
W_+(t)&=&\frac{i}\hbar g\int_\infty^t dt'  \ \langle 1,f|A(t) |0,i\rangle \\ \nonumber 
&=&\frac{i}\hbar g\int_\infty^t dt' \ e^{i(t-t')\epsilon}  \ \langle f|e^{-iHt}A(t') |i\rangle.
\end{eqnarray}
The probability is the square of the amplitude summed over the final states, which gives
$$
P_+(t)\! =\!\frac{g^2}{\hbar^2}\! \int_\infty^t \! \!\! \! dt'\! \!\int_{-\infty}^t\! \! \!\!\!\!\!\! dt'' \ e^{i(t'-t'')\epsilon}  \ \langle i|A(t')A(t'') |i\rangle. 
$$
Assuming time independence, the integrand depends only on the difference $s=t''-t'$ 
$$
P_+(t)=g^2 \int_\infty^t dT\int_{-\infty}^T ds \ e^{is\epsilon}  \ \langle i|A(s)A(0) |i\rangle. 
$$
and the probability to move up per unit time is therefore, in the large-time limit,
$$
p_+=\frac{P_+(t)}{dt}=g^2 \int_{-\infty}^\infty dt \ e^{is\epsilon}  \ \langle i|A(s)A(0) |i\rangle,
$$
namely the value in $\epsilon$ of the Fourier transform  $\tilde f_{AA}$ of 
\be
f_{AA}(t)\equiv\langle i|A(t)A(0) |i\rangle. 
\ee
Repeating the same calculation for the probability to go down, we obtain the value of $\tilde f_{AA}$ in $-\epsilon$. Now, if the interaction does not disturb the system much and if the ratio of the two happens to be 
\be
\frac{p_+}{p_-}=e^{-\beta\epsilon},
\ee
then the equilibrium distribution for the detector will be Boltzmannian, at inverse temperature $\beta$.  Therefore we find that the condition for the thermometer to measure the inverse temperature $\beta$ is that 
\be
\tilde f_{AA}(-\epsilon)=e^{-\beta\epsilon}\ \tilde f_{AA}(\epsilon).
\ee
In Fourier transform, this reads 
\be
 f_{AA}(-t)=  f_{AA}(t-i\beta).
\ee
which can be recognised as a form of the celebrated KMS condition, which characterises equilibrium \cite{Haag:1992hx}.  

Armed with this general tool, it is now easy to derive the Unruh effect. If the detector moves in a quantum field in the vacuum state, following a trajectory $x(s)$, where $s$ is the proper time along the trajectory, and $A$ is the field operator, then 
\begin{eqnarray}
f_{AA}(s)&=&\langle i|A(s)A(0) |i\rangle  \nonumber \\
&=&\langle 0|\phi(x(s))\phi(x(0)) |0\rangle
\end{eqnarray}
is simply the two-point function of the field theory, along the trajectory. The trajectory of a detector moving at constant acceleration in Minkowski space is 
\be
x(s)=\left(\frac1a\sinh(as),\frac1a\cosh(as), 0 , 0\right).  \label{tra}
\ee
Therefore it follows from the previous general discussion that a constantly accelerated detector sees a temperature $\beta$ if the two-point function of the quantum field along the trajectory 
\be
f(s)=\langle 0|\phi(x(s))\phi(x(0)) |0\rangle
\ee
satisfies the KMS condition 
\be
 f(s)=  f(-s+i\beta).   \label{KMS}
\ee
or equivalently, in Fourier transform,
\be
\tilde f(-\epsilon)=e^{-\beta\epsilon}\ \tilde f(\epsilon).   
\ee
Observe that this is a condition on the Fourier components of the two point function at the energy scale at which the thermometer works. Short distance physics plays no role here. 

For a free massless field, the two point function is proportional to the square of the inverse of the 4-distance.  
\be
\langle 0|\phi(x)\phi(0)) |0\rangle\sim\frac1{|x|^2}
\ee
 Along the trajectory:
\be
f(s)=\langle 0|\phi(x(s))\phi(x(0))) |0\rangle\sim\frac1{|x(s)-x(0)|^2}
\ee
This can be easily computed from (\ref{tra}), giving
\be
f(s)=\frac2{a^2}\left(\cosh(as)-1\right),
\ee
which satisfies the KMS condition (\ref{KMS}) with the Unruh temperature $a/2\pi$.  This proves that an accelerated detector in the vacuum of a free massless theory measures the Unruh temperature. 

In LQC the two-point function at the scale of the acceleration is not affected by quantum gravitational corrections, since these become relevant only at the Planck scale.  In particular, the two-point function for the perturbative excitations of the gravitational field has been computed in covariant quantum gravity for the Lorentzian theory in  \cite{Bianchi:2011hp,Han:2011re,Han:2013tap}. The result of these works is that at the lowest order the two-point function converges to the free one, in the large distance limit.  Here ``large" means large with respect to the Planck scale.   Therefore it is clear that LQG has no effect whatsoever on the prediction of the Unruh effect. 

What therefore went wrong in \cite{Hossain2014}?  The answer is interesting.  As observed by Sabine Hossenfelder \cite{Hossenfelder}, trying to compute a quantity in a finite theory like LQG by means of an infinite renormalisation is, besides being perverse, also a doubtful procedure. One risks to subtract ``one infinity too much"  \cite{Hossenfelder}.  By complicating a simple low-energy effect, expressing it in terms of the difference between divergent sums, one gets things wrong.    

This is probably a general lesson in quantum gravity. The theory is finite, because of the Planck scale cut off provided by the physical discreteness of the geometry. Infinite renormalisation calculation, in this context, obscure, rather than clarifying, the physics. 

\providecommand{\href}[2]{#2}\begingroup\raggedright\endgroup


\end{document}